\documentclass[twocolumn, dvipdfmx]{aastex701} 
\usepackage{amsmath,amsfonts,amssymb}
\usepackage{bm}
\usepackage{graphics}
\usepackage{graphicx}
\usepackage{here} 
\usepackage{type1cm}
\usepackage{multirow}
\usepackage{float}
\usepackage{natbib}
\usepackage{comment}

\def\kms{\hbox{km s$^{-1}$}}
\def\VLSR{\hbox{$V_{\rm LSR}$}}

\def\Hone{\hbox{H\,{\scriptsize I}}}

\shorttitle{Molecular Cloud toward Fermi J1913+0515}
\shortauthors{Ariyama et al.}

\begin{document}

\title{Detection of a Molecular Cloud toward the Heartbeating Gamma-ray Source near the Microquasar SS 433}


\author[orcid=0000-0002-5566-0634,gname=Tomoharu,sname=Oka]{Tomoharu Oka} 
\affiliation{School of Fundamental Science and Technology, Graduate School of Science and Technology, Keio University, 3-14-1 Hiyoshi, Kohoku-ku, Yokohama, Kanagawa
223-8522, Japan}
\affiliation{Department of Physics, Institute of Science and Technology, Keio University, 3-14-1 Hiyoshi, Kohoku-ku, Yokohama, Kanagawa 223-8522, Japan}
\email{tomo@phys.keio.ac.jp}

\author[gname=Ryo, sname=Ariyama]{Ryo Ariyama}
\affiliation{Department of Physics, Institute of Science and Technology, Keio University, 3-14-1 Hiyoshi, Kohoku-ku, Yokohama, Kanagawa 223-8522, Japan}
\affiliation{School of Fundamental Science and Technology, Graduate School of Science and Technology, Keio University, 3-14-1 Hiyoshi, Kohoku-ku, Yokohama, Kanagawa
223-8522, Japan}
\email{sirius@keio.jp}

\author[orcid=0009-0006-9842-4830,gname=Tatsuya,sname=Kotani]{Tatsuya Kotani}
\affiliation{Department of Physics, Institute of Science and Technology, Keio University, 3-14-1 Hiyoshi, Kohoku-ku, Yokohama, Kanagawa 223-8522, Japan}
\affiliation{School of Fundamental Science and Technology, Graduate School of Science and Technology, Keio University, 3-14-1 Hiyoshi, Kohoku-ku, Yokohama, Kanagawa
223-8522, Japan}
\email{sci.tatsu.729@keio.jp}

\begin{abstract}
We report the detection of a molecular cloud, CO+40.05--2.40, positionally coincident with the ``heartbeating'' GeV source Fermi J1913+0515 at the northern boundary of the SS 433/W50 system. Millimeter and submillimeter spectroscopy with the Nobeyama 45 m telescope and the James Clerk Maxwell Telescope shows that the cloud has physical properties typical of quiescent dark clouds in the Galactic disk, with no evidence of shock heating or enhanced excitation. We examine possible high-energy emission mechanisms and find that the observed GeV luminosity cannot be accounted for by electron bremsstrahlung or hadronic interactions driven by relativistic particles originating from SS 433 under reasonable energetic assumptions. As an alternative, we propose that the $\gamma$-rays may arise from a compact object embedded within the cloud and powered by Bondi-type accretion. In this framework, the reported heartbeat-like variability may reflect periodic modulation of the accretion flow by density waves induced by the precessing equatorial outflow of SS 433.
\end{abstract}

\section{Introduction}\label{sec:intro}
SS 433, the first microquasar discovered in our Galaxy, is situated approximately 5.5 kpc from the Solar System \citep{GaiaCol_2016, Blundell_Bowler_2004}, at the center of the supernova remnant W50 nebula. This object is characterized by highly collimated bipolar jets of plasma ejected at a velocity of $0.26 c$ in the polar direction \citep{Fabian_Rees_1979, Fabrika_2004}. The extension of these jets, known as jet lobes, are observed to stretch the surrounding W50 nebula in a bipolar manner \citep{Dubner_1998}. Furthermore, the SS 433 jets precess with an opening angle of approximately $20\arcdeg$ and a period of 162.250 days \citep{Davydov_2008, Blundell_Bowler_2004}. In addition to the bipolar jets, an equatorial outflow, often referred to as a disk wind, perpendicular to the jets has also been observed from SS 433 \citep{Margon_Anderson_1989}.  While the velocity of this equatorial outflow is significantly lower, estimated at $500\mbox{--}1000$ \kms\ \citep{Blundell_2011} (approximately 2.5 orders of magnitude lower than that of the bipolar jets), its mass-loss rate is considerably higher \citep{Fuchs_2002}.  Resultantly, both the bipolar jets and the equatorial outflow from SS 433 are found to be ejected with comparable kinetic powers of $10^{39\mbox{--}40}$ erg s$^{-1}$ \citep{Fabrika_2004, Middleton_2021}.

Within the SS 433/W50 system, the GeV gamma-ray source Fermi J1913+0515 has been detected. This source, cataloged in the Fermi-LAT 8-year Point Source List, emits in the 100 MeV--300 GeV energy band and stands out as the brightest GeV gamma-ray source in the region \citep{Fang_2020, Li_2020}. It is located at the northern edge of the W50 radio shell, approximately $0.4\arcdeg$ ($\sim$40 pc at a distance of 5.5 kpc) away from SS 433. Notably, this position is neither aligned with the extension of the jet lobes nor falls within the precessional cone defined by the extended ejection direction of the jets. Despite this spatial discrepancy, the gamma-ray flux of Fermi J1913+0515 was surprisingly reported to blink with a period of $160.88\pm2.66$ days \citep{Li_2020}. This period is remarkably close to the SS 433's precessional period (162.250 days), strongly implying a direct physical link.

\begin{figure*}[htbp]
\begin{center}
\includegraphics[clip, width=0.8\textwidth]{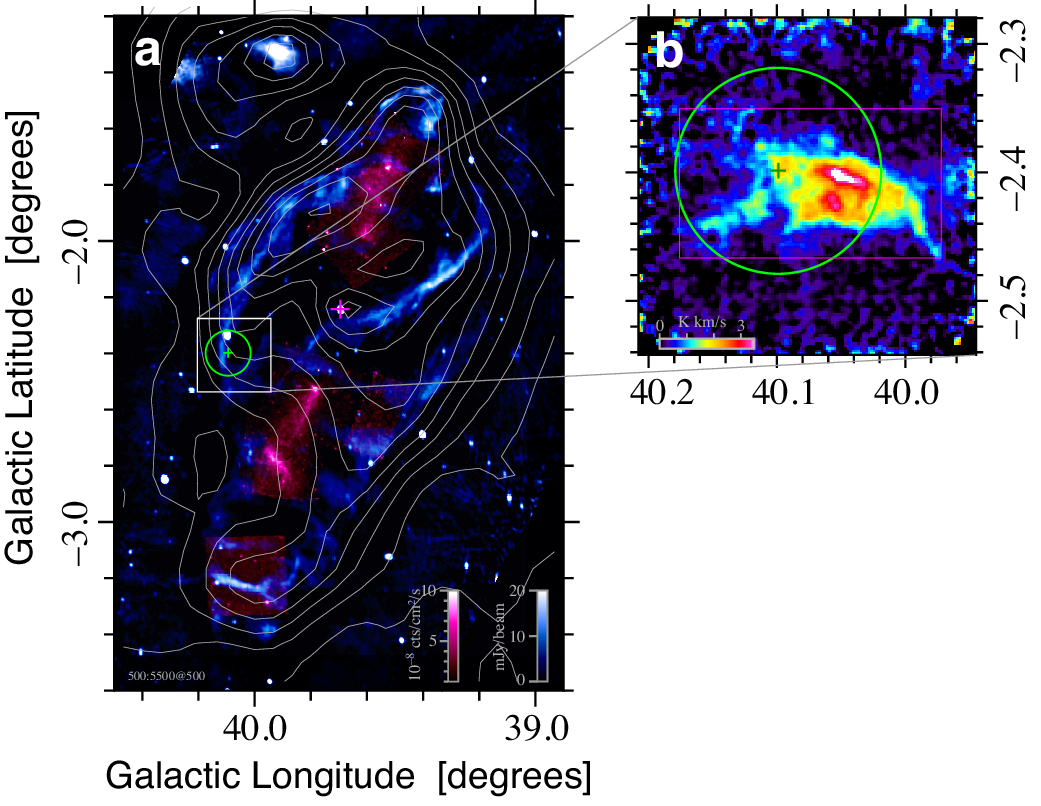}
\caption{(a) Radio continuum and hard X-ray images of the SS 433/W50 system.  The black-blue-white colormap shows the JVLA L-band image \citep{Sakemi_2021}, while the black-magenta-white colormap represents the {\it Chandra} $0.5\mbox{--}7$ keV image \citep{Tsuji_2025}.  Gray contours indicate the 1.4 GHz continuum flux observed with the Effelsberg 100 m telescope \citep{Reich_1982,Reich_1986}.  The magenta cross marks the position of SS 433.  The green circle denotes the 95\%\ positional error circle of Fermi J1913+0515, and the white square outlines the area shown in panel (b).  (b) Velocity-integrated map of CO {\it J}=3--2 emission obtained with ASTE.  The green circle is the same as in panel (a), and the magenta rectangle denotes the mapping area covered by the NRO 45m observations in 2023. }
\label{fig:guide}
\end{center}
\end{figure*}

A clear understanding of the gamma-ray emission process and the origin of Fermi J1913+0515's periodicity remains elusive. While an HI cloud is spatially adjacent to the source \citep{Li_2020}, and a scenario proposes its irradiation by SS 433's equatorial outflow causes gamma-ray emission via neutral pion decay, this hypothesis faces challenges. The center of this HI cloud is located outside the error circle of Fermi J1913+0515, and its size cannot explain the gamma-ray `heartbeating' on a $\sim$half-year timescale.  

In this paper, we report the detection of a molecular cloud (CO+40.05--2.40) located precisely within the error circle of Fermi J1913+0515. This cloud represents a more promising clump of baryonic matter that could contribute to the gamma-ray emission from Fermi J1913+0515, compared to the aforementioned HI cloud.

\section{Observations} \label{sec:obs}
\subsection{NRO 45m Observations} \label{subsec:nro}
We observed the {\it J}=1--0 transitions of CO (115.271 GHz), $^{13}$CO (110.201 GHz), C$^{18}$O (109.782 GHz), N$_{2}$H$^{+}$ (93.174 GHz), HCO$^{+}$ (89.189 GHz), and HCN (88.632 GHz) with the Nobeyama Radio Observatory (NRO) 45 m telescope. Initial CO {\it J}=1--0 observations were performed on February 22, 2022, covering a $12\arcmin$ square region that encompassed the error circle of Fermi J1913+0515. After confirming the spatial extent of the CO emission, we re-mapped a $740\arcsec\!\times\!420\arcsec$ rectangular region (Fig.\ref{fig:guide}) covering the full extent of the CO lines between January and February 2023. Additionally, we obtained N$_{2}$H$^{+}$, HCO$^{+}$ and HCN maps for a $110\arcsec\!\times\!110\arcsec$ region centered on the densest emission area between February and March 2023.  

We used the four-beam receiver system FOREST \citep{Minamidani_2016} on the 45-m telescope in conjunction with the SAM45 spectrometer \citep{Kuno_2011, Kamazaki_2012}. The half-power beamwidths (HPBWs) were approximately $14\arcsec$ at 110 GHz and $19\arcsec$ at 86 GHz. The SAM45 was operated in 250 MHz bandwidth mode with a frequency resolution of 122.07 kHz, providing a velocity coverage of $681\,\kms$ and a velocity resolution of $0.333\,\kms$ at $110$ GHz.  The system noise temperature ($T_{\rm sys}$) typically ranged from 120 to 400 K during the CO observations, and from 100 to 150 K for the N$_{2}$H$^{+}$, HCO$^{+}$ and HCN observations.  The pointing accuracy of the telescope was checked and corrected every $1.5\mbox{--}2$ hours by observing the SiO maser source R-Aql at 43 GHz and was maintained within $3\arcsec$ in both azimuth and elevation.  

Antenna temperature ($T_{\rm A}^*$) calibration was performed using the standard chopper-wheel method. We also observed W51A on a daily basis as a reference source for intensity calibration, from which we derived the main-beam efficiency ($\eta_{\rm MB}$) for each beam.  The antenna temperature was then converted to the main-beam temperature ($T_{\rm MB}$) by multiplying by $1/\eta_{\rm MB}$. The maps were convolved using the Bessel--Gaussian function and resampled onto a $7\farcs 2\!\times\!7\farcs 2\!\times\!0\fdg 25\,\kms$ regular grid. The resultant rms noise levels are $\Delta T_{\rm MB} = 0.12\,{\rm K}$ for the CO maps and $0.02\,{\rm K}$ for the N$_{2}$H$^{+}$ and HCN maps.

\begin{figure*}[htbp]
\begin{center}
\includegraphics[clip, width=1.0\textwidth]{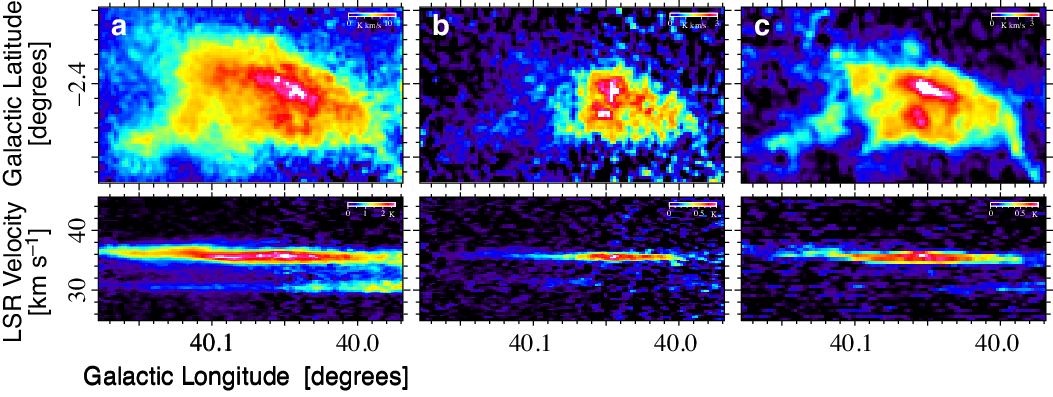}
\caption{(Top) Maps of velocity-integrated emission of (a) CO {\it J}=1--0, (b) $^{13}$CO {\it J}=1--0, and (c) CO {\it J}=3--2 lines.  The integration velocity range is $\VLSR\!=\!34\sim38\,\kms$.  (Bottom) Longitude velocity maps of (a) CO {\it J}=1--0, (b) $^{13}$CO {\it J}=1--0, and (c) CO {\it J}=3--2 lines.  The integration latitude range is $b\!=\!-2\fdg 46\sim -2\fdg 35$. 
} 
\label{fig:composite}
\end{center}
\end{figure*}

\subsection{JCMT Observations} \label{subsec:jcmt}
CO {\it J}=3--2 (345.796 GHz) and HCN {\it J}=4--3 (354.506 GHz) lines were observed with the James Clerk Maxwell Telescope (JCMT) between August and September 2022. We mapped a $720\arcsec$ square region centered at the peak of CO {\it J}=1--0 integrated intensity, $(l, b)\!=\!(40\arcdeg 04\arcmin 12\farcs 0, -02\arcdeg 24\arcmin 36\fdg 0)$. We used the 345 GHz band receiver HARP \citep{Buckle_2009} with the ACSIS spectrometer.  The HPBW was approximately $14\arcsec$ at 345 GHz. The ACSIS was operated at in 400 MHz bandwidth mode with a frequency resolution of 61.0 kHz, providing a velocity coverage of $348\,\kms$ and a velocity resolution of $0.053\,\kms$ at $345$ GHz. The $T_{\rm sys}$ typically ranged from 300 to 500 K during the observations. 

Antenna temperature was calibrated using the standard chopper-wheel method. We adopted the main beam efficiency value published by the observatory, $\eta_{\rm MB}\!=\!0.64$. The maps were resampled onto a $7\farcs 2\!\times\!7\farcs 2\!\times\!0\fdg 25\,\kms$ regular grid, and the resultant rms noise level is $\Delta T_{\rm MB} = 0.1\,{\rm K}$.

\section{Results} \label{sec:results}
\subsection{Detection of CO+40.05-2.40} \label{subsec:detection}
Our observations have revealed a molecular cloud, designated CO+40.05--2.40, which lies largely within the 95\%\ positional error circle of Fermi J1913+0515 (Figure \ref{fig:guide}b).  The cloud spans an angular size of approximately $0\fdg 1\!\times\!0\fdg 05$ and exhibits a west-to-east trailing morphology, with peak intensity concentrated near its geometric center.

As shown in Figure \ref{fig:composite}, the cloud consists of a spatially extended envelope traced by the low-excitation CO {\it J}=1--0 line, a more compact and westward-biased component traced by the higher-excitation CO {\it J}=3--2 line, and a centrally concentrated high-column-density core evident in $^{13}$CO {\it J}=1--0.  The cloud is detected within a velocity range of $\VLSR\!=\!33$ to $38\,\kms$, with a central velocity of $\VLSR\!=\!35.5\,\kms$.  Within the high-column-density dense core, emission lines of C$^{18}$O {\it J}=1--0, N$_2$H$^{+}$ {\it J}=1--0, HCO$^{+}$ {\it J}=1--0, and HCN {\it J}=1--0 were detected, whereas HCN {\it J}=4--3 was not.  Emission from high-density tracers (HCN, HCO$^{+}$, and N$_2$H$^{+}$) also confirms the presence of dense ($n({\rm H}_2)\!\gtrsim\! 10^5\,{\rm cm}^{-3}$) gas.  Because the results of the high-density probes are not essential to the discussion that follows, we do not consider them further in this paper.

\subsection{Association with the SS 433/W50 System}
The central velocity of CO+40.05--2.40, $\VLSR\!=\!35.5\,\kms$, corresponds to a kinematic distance of either $2.0$ or $11.0\,\mathrm{kpc}$, assuming a Galactic rotation curve \citep{Clemens_1985}.  Adopting the nearer value places the cloud at a distance of approximately half that to SS 433 ($5.5\,\mathrm{kpc}$). However, this kinematic estimate should be treated with caution. The SS 433/W50 system lies near the end of the Galactic bar, where non-circular motions may significantly affect radial velocities. 

\Hone 21 cm spectral line data reveal an expanding cavity centered at $\VLSR\!\simeq\!42\,\kms$, spatially coincident with the W50 radio shell \citep{HI4PI_2016}. The systemic velocity and independently determined distance of SS~433 are consistent with gas at $V_{\rm LSR}\sim40\,\kms$ in this direction when such non-circular motions are considered. The velocity offset between the cloud ($35.5\,\kms$) and the cavity center ($\sim\! 42\,\kms$) is modest and comparable to the internal velocity dispersion of large-scale structures in this region.  CO+40.05--2.40 lies along the edge of this \Hone\ cavity in position--velocity space.

Taken together, these spatial and kinematic consistencies strongly support a physical association of the \Hone\ cavity with the SS 433/W50 system rather than a chance superposition along the line of sight. Since CO+40.05--2.40 traces the boundary of this structure, we consider it likely to belong to the same physical environment and adopt a distance of $5.5\,\mathrm{kpc}$ in the following discussion.

\subsection{Physical Parameters} \label{subsec:physical_params}
Based on the observed CO emissions, we derived the physical parameters of CO+40.05--2.40.  Following the definition by \citet{Solomon_1987}, the size parameter $S$, velocity dispersion $\sigma_{\mathrm{V}}$, and CO luminosity $L_{\mathrm{CO}}$ were estimated from the CO {\it J}=1--0 data to be $S\!=\!3.8\,\mathrm{pc}$,  $\sigma_\mathrm{V}\!=\!1.1\,\kms$, and $L_{\mathrm{CO}}\!=\!10^{3.1}\,\mathrm{K\,km\,s^{-1}\,pc^2}$, respectively. Using these values, the virial theorem mass of the cloud is calculated to be $M_{\mathrm{VT}}\!=\!10^{4.0}\,M_{\sun}$.  These $L_{\mathrm{CO}}$ and $M_{\mathrm{VT}}$ values closely follow the $L_{\mathrm{CO}}\mbox{--}M_{\mathrm{VT}}$ relation for Galactic disk clouds \citep{Solomon_1987}.

Next, we estimate the LTE mass of CO+40.05--2.40 from the CO line intensities.  At the $^{13}$CO {\it J}=1--0 intensity peak, located at $(l, b, V_{\mathrm{LSR}}) = (40\fdg04, -2\fdg40, 35.6\,\kms)$, we derived the optical depth $\tau_{12}\!=\!36$ and the excitation temperature $T_{\mathrm{ex}}\!=\!9.8\,\mathrm{K}$ from the main-beam temperatures of the CO and $^{13}$CO lines. Here, we assumed a carbon isotope ratio $[^{12}\mathrm{C}]/[^{13}\mathrm{C}]\!=\!60$ \citep{Henkel_1982} and a carbon monoxide abundance relative to H$_2$ of [CO]/[H$_2$]$\!=\! 10^{-4.1}$ \citep{Frerking_1982}, both of which are typical values for the Galactic disk.  Using the derived excitation temperature and the integrated intensity of the $^{13}$CO line, we obtained an LTE mass of $M_{\mathrm{LTE}}\!=\!10^{3.2}\,M_{\sun}$ in the optically thin limit.  This method is regarded as adequate to first order, as $\tau_{13}(\!=\!\tau_{12}/60)\!=\!0.6$ at the $^{13}$CO peak, suggesting that the optical depth is lower over most of the cloud.

\textbf {The LTE mass derived here is about one order of magnitude smaller than the virial mass. This difference may reflect departures from simple virial equilibrium, since non-gravitational contributions such as turbulence or external pressure can broaden the observed line width and lead to an overestimate of $M_{\rm VT}$. Conversely, the LTE mass may represent a lower limit depending on the adopted excitation and optical depth assumptions. We therefore regard the two estimates as bracketing the plausible mass range of the cloud.}

The integrated intensity ratio of CO {\it J}=3--2 to {\it J}=1--0 is 0.3, a value typical for molecular clouds in the Galactic disk.  No significant enhancement of the CO {\it J}=3--2/{\it J}=1--0 ratio is found in any spatial or velocity domain of CO+40.05--2.40. The mean H$_2$ density is estimated to be $\langle n_{\mathrm{H}_2}\rangle\!\sim\!10^{2.0}\,\mathrm{cm}^{-3}$, based on the LTE-derived H$_2$ column density toward the $^{13}$CO intensity peak ($N_{\mathrm{H}_2}\!=\!10^{21.4}\,\mathrm{cm}^{-2}$) and by assuming a line-of-sight depth of $2S$. An LVG analysis of the CO {\it J}=1--0, 2--1, and 3--2 main-beam temperatures toward the same position yields loose constraints on the physical conditions, namely $n_{\mathrm{H}_2}\!\lesssim\!10^{3.2}\,\mathrm{cm}^{-3}$, $N_{\mathrm{H}_2}/dV\!\gtrsim\!10^{21.5}\,\mathrm{cm}^{-2} (\kms)^{-1}$, and $T_{\rm k}=9.0\mbox{--}10.3\,\mathrm{K}$ at the $95$\% confidence level. These values are consistent with the LTE-derived estimates.

\begin{deluxetable*}{lccc}
\tablecaption{Adopted Parameters for CO+40.05--2.40 and Fermi~J1913+0515 \label{tab:params}}
\tablehead{
\colhead{Quantity} & \colhead{Symbol} & \colhead{Adopted value} & \colhead{Note}
}
\startdata
Distance to SS 433/W50       & $d$                   & $5.5$ kpc                       & Association with W50 \\
Angular size of CO+40.05--2.40 & $\Delta l \times \Delta b$ & $0\fdg1 \times 0\fdg05$         & From CO $J=3$--2 map \\
Cloud radius                  & $S$                   & $3.8$ pc                        & At $d = 5.5$ kpc \\
LTE mass                      & $M_{\rm LTE}$        & $10^{3.2}\,M_\odot$             & This work \\
Mean hydrogen density         & $\langle n_{\rm H}\rangle$ & $10^{2.5}\,{\rm cm^{-3}}$       & From $M_{\rm LTE}$ and $S$ \\
Total number of target protons & $N_{\rm H}$         & $10^{60.3}$                     & Derived from $M_{\rm LTE}$ \\
$\gamma$-ray luminosity       & $L_\gamma$           & $10^{34.6}\,{\rm erg\,s^{-1}}$  & $0.1\mbox{--}300$ GeV, $d=5.5$ kpc \\
Effective bremsstrahlung cross section & $\langle\sigma_{\rm br}\rangle$ & $(1\text{--}3)\!\times\!10^{-26}\,{\rm cm^{2}}$ & Literature value \\
Mean photon energy & $\langle E_{\gamma}\rangle$ & $(0.2\text{--}0.8)\,{\rm GeV}$  &  $0.1\mbox{--}300$ GeV, $p\!=\!2.0\text{--}3.0$ \\
Effective p--p cross section &  $\langle\sigma_{pp,\gamma}k_{\gamma}\rangle$  &  $(0.5\mbox{--}1)\!\times\!10^{-26}\,{\rm cm^{2}}$ & Literature value 
\enddata
\tablecomments{These parameters are adopted throughout Section~\ref{sec:discussion} for estimating the $\gamma$-ray production efficiency in the leptonic, hadronic, and accretion scenarios.}
\end{deluxetable*}

\section{Discussion} \label{sec:discussion}
\subsection{Physical Relationship with Fermi J1913+0515} \label{subsec:relationship}
Molecular clouds are widely recognized as efficient $\gamma$-ray emitters when interacting with high-energy cosmic rays, either through hadronic $\pi^{0}$ decay or relativistic bremsstrahlung processes \citep[e.g.,][]{Aharonian_1994, Gabici_2009}.  Well-studied examples include IC 443, where shocked molecular gas has been firmly associated with cosmic rays accelerated in the supernova remnant \citep{Abdo_2010}, and W28 and W44, in which GeV--TeV $\gamma$-rays arise as cosmic rays from adjacent remnants collide with dense molecular material \citep{Giuliani_2010, Ackermann_2013}.  Possible associations between unidentified $\gamma$-ray sources and molecular clouds have also been suggested; for instance, \citet{Oka_1999} reported a candidate case involving a rotation-powered pulsar and a nearby molecular cloud.  Together, these systems demonstrate that dense molecular gas, when bombarded by relativistic protons and electrons from nearby accelerators, provides abundant target material capable of producing $\gamma$-rays through neutral pion decay and electron bremsstrahlung.

CO+40.05--2.40 lies predominantly within the 95\% positional error circle of Fermi J1913+0515.  Using the CfA CO survey data ($0\fdg 125$ grid; \citealt{Dame_2001}), we find that the area where the CO integrated intensity exceeds $5\,\mathrm{K}\,\kms$ in the velocity range of CO+40.05--2.40 ($V_{\mathrm{LSR}} = 34\mbox{--}38\,\kms$) occupies only $0.094\,\mathrm{deg}^{2}$ within a $2\fdg 125\!\times\!2\fdg 125$ region centered on the error circle.  The chance probability that a cloud comparable to CO+40.05--2.40 would lie inside the error circle (area $\simeq\!0.02\,\mathrm{deg}^2$) is only about $2.7\%$.  This low probability strongly suggests that the alignment is physical rather than coincidental.

\subsection{Gamma-ray Emission from CO+40.05-2.40} \label{subsec:gamma_emission}
The observed $\gamma$-ray flux of Fermi J1913+0515 in the $0.1\mbox{--}300$\,GeV band is $F_{\gamma}\!\simeq\! 10^{-10.9}\,{\rm erg\,cm^{-2}\,s^{-1}}$ \citep{Li_2020}, which corresponds to a luminosity of $L_{\gamma}\!\simeq\! 10^{34.6}\,{\rm erg\,s^{-1}}$ for a distance of $5.5$ kpc.  If the two are physically associated, CO+40.05--2.40 would represent a dense baryonic target.  Its LTE mass ($10^{3.2}\,M_{\sun}$) implies a proton content of $N_{\rm H}\!\sim\! 10^{60.3}$, and assuming a spherical radius of $S=\!3.8$ pc yields a mean density $\langle n_{\rm H}\rangle\!\sim\!10^{2.5}\,{\rm cm^{-3}}$.  Using these physical quantities (summarized in Table \ref{tab:params}), we evaluate three possible mechanisms --- leptonic bremsstrahlung, hadronic proton--gas interactions, and accretion-powered emission.  

\vspace{3pt}
\subsubsection{Leptonic Scenario} \label{subsec:electronic}
The total rate of photon production by electron bremsstrahlung is
\begin{equation}
\dot{N}_{\gamma}
= N_{\rm H}\, \bar{\Phi}_{e}\, \langle \sigma_{\rm br} \rangle ,
\label{eq:Ngamma_def}
\end{equation}
and the resulting energy luminosity is  
\begin{equation}
L_{\gamma} \simeq \dot{N}_{\gamma}\, \langle E_{\gamma} \rangle
= N_{\rm H}\, \bar{\Phi}_{e}\, \langle \sigma_{\rm br} \rangle\, \langle E_{\gamma} \rangle .
\label{eq:Lgamma_from_Ngamma}
\end{equation}
Here $\langle\sigma_{\rm br}\rangle$ denotes the effective, band-integrated bremsstrahlung cross section for producing photons in the $0.1\mbox{--}300$ GeV range, and $\langle E_\gamma\rangle$ is the corresponding mean photon energy for a power-law electron spectrum with index $p\!=\!2.0\mbox{--}3.0$ (see Table \ref{tab:params}).  Solving for the required electron flux yields $\bar{\Phi}_{e}\!\sim\!10^{2.7\text{--}3.8}\,{\rm cm^{-2}\,s^{-1}}$, which is a value several orders of magnitude above the local Galactic cosmic-ray electron flux.  For a spherical cloud with a radius of $S$, the total electron injection rate is $\dot{N}_{e}\!\simeq\!(0.2\text{--}2.7)\!\times\!10^{42}\,{\rm s^{-1}}$ and the corresponding energy deposition rate is $\dot{E}_{e}\!\sim\!10^{37.8\text{--}39.5}\,{\rm erg\,s^{-1}}$.  

Although the required electron power is formally comparable to the mechanical luminosity of SS 433 jets and disk wind ($\sim\!10^{38\text{--}39}\,{\rm erg\,s^{-1}}$), only a very small fraction of the outflow can geometrically intercept the cloud.  Given that CO+40.05--2.40 subtends only $\Delta\Omega/4\pi\!\sim\!10^{-2.6}$ as viewed from SS 433, the power available to irradiate the cloud would be reduced to at most $\sim\!10^{35.4\text{--}36.4}\,{\rm erg\,s^{-1}}$, far below the electron power that must be supplied.  This geometric constraint therefore renders the leptonic scenario robustly disfavored.

\subsubsection{Hadronic Scenario} \label{subsec:hadronic}
For proton--proton interactions, the corresponding expression is
\begin{equation}
L_{\gamma} \simeq N_{\rm H}\,\bar{\Phi}_{p}\,\langle\sigma_{pp,\gamma}k_{\gamma}\rangle \, \langle E_{\gamma} \rangle ,
\end{equation}
where $k_{\gamma}$ denotes the $\gamma$-ray multiplicity per proton-proton collision.  Here, $\bar{\Phi}_{p}$ is the confinement-enhanced effective particle flux, reflecting the fact that relativistic protons can undergo repeated scattering and remain trapped within the cloud rather than interacting only once.  Adopting $\langle\sigma_{pp,\gamma}k_{\gamma}\rangle\!\simeq\! (0.5\mbox{--}1)\!\times\!10^{-26}\,{\rm cm^{2}}$, as representative for a proton power-law spectrum with index $p\!=\!2\mbox{--}3$, we obtain the required effective proton flux $\bar{\Phi}_{p}\!\sim\!10^{3.2\mbox{--}4.1}\,{\rm cm^{-2}\,s^{-1}}$.  However, this value does not represent the incoming flux from outside of the cloud, because protons interacting inside the cloud suffer multiple scatterings and are at least partially confined by the magnetic field of order a few $\mu$G.  

Unlike relativistic electrons, GeV protons suffer negligible radiative losses in molecular cloud environments and can remain confined for timescales longer than the dynamical crossing time. Electrons of comparable rigidity cool more rapidly, preventing efficient accumulation. The confinement enhancement considered here is thus primarily relevant to the hadronic case.

Under standard diffusion-confinement arguments for dense molecular gas, the injection flux is reduced approximately by the enhancement factor $f_{\rm conf}\!=\! t_{\rm conf}/t_{\rm cross}\!\sim\! 10^{2.8}$, implying $\bar{\Phi}_{p}^{\rm (inj)}\!\sim\!10^{0.4\mbox{--}1.3}\,{\rm cm^{-2}\,s^{-1}}$.  For a cloud radius of $S$, the corresponding proton injection rate becomes $\dot{N}_{p}\!\simeq\!(1.1\text{--}9.3)\!\times\!10^{39}\,{\rm s^{-1}}$.  Adopting a representative mean proton energy $\langle E_{p}\rangle\!\sim\! 10\langle E_{\gamma}\rangle$, the required proton power is $\dot{E}_{p}\!\sim\!10^{36.5\text{--}38.0}\,{\rm erg\,s^{-1}}$.  This still exceeds the mechanical power available from SS 433 within the solid angle subtended by the cloud, $L_{\rm mech}^{\Omega}\!\sim\! 10^{35.4\mbox{--}36.4}\,{\rm erg\,s^{-1}}$.  

Although energetically less extreme than the leptonic case, the hadronic scenario remains insufficient unless the proton injection is confined to a narrowly restricted solid angle.  Even when adopting optimistic values for the effective cross section, mean proton energy, and confinement enhancement within the cloud,
the required proton power remains comparable to or exceeds the mechanical energy available from SS 433.

\subsubsection{Accretion Scenario} \label{subsec:accretion}
An alternative is that the $\gamma$-rays originate not from external irradiation but from a compact object embedded within CO+40.05--2.40.  A neutron star or black hole accreting molecular gas at the Bondi rate, 
\begin{equation}
\dot{M}_{\rm Bondi}
\simeq 4\pi G^{2} M^{2} \rho\, c_{\rm s}^{-3},
\end{equation}
would yield an accretion power  
\begin{equation}
L_{\rm acc} \simeq \eta\, \dot{M}_{\rm acc} \, c^{2} .
\end{equation}
For $T\!\sim\!10\, {\rm K}$ and  a $\sim\!10\,M_{\odot}$ compact object, we obtain $\dot{M}_{\rm Bondi}\!\sim\!10^{-11}\mbox{--}10^{-10}\,M_{\odot}\,{\rm yr^{-1}}$, corresponding to $L_{\rm acc}\!\sim\!10^{34\mbox{--}35}\,{\rm erg\,s^{-1}}$ for radiative efficiencies $\eta \sim 10^{-3}\mbox{--}10^{-1}$.  

This level is comparable to the inferred luminosity of Fermi J1913+0515, implying that an accreting compact object could, in principle, power the emission.  However, the absence of detectable radio, IR, or X-ray emission implies either a radiatively inefficient accretion state (e.g., RIAF/ADAF) or an unusually high conversion efficiency to $\gamma$-rays---conditions not yet confirmed in known systems.  Such a $\gamma$-ray--dominated accretion state has not yet been confirmed observationally, but shares phenomenological similarities with transitional millisecond pulsars \citep[e.g., PSR J1023+0038;][]{Papitto_2015,Deller_2015,Baglio_2023} and ultra-low luminosity accreting black hole systems \citep[e.g., 3FGL J1544.6--1125;][]{Jaodand_2021}, suggesting that it may occupy an unexplored region of the accretion-mode parameter space.

The apparent spatial coincidence between a compact object and a dense molecular cloud may seem unlikely at first glance. However, stellar-mass compact objects are expected to be abundant in the Galactic disk, and their typical space velocities (of order a few $\times 10\,\kms$) do not preclude temporary residence within molecular gas structures. If the relative velocity between the object and the ambient gas is sufficiently small, Bondi-type accretion can proceed efficiently without requiring fine tuning. In this context, the scenario does not require an extraordinary coincidence, although direct observational confirmation of such an embedded object remains essential. Moreover, the lifetime of molecular clouds ($10^{6}$--$10^{7}$ yr) is long compared to the dynamical crossing time of compact objects through the Galactic disk, increasing the probability of transient encounters. Observational precedents for such configurations have been reported; for example, \citet{Oka_1999}  identified a pulsar wind nebula spatially associated with a molecular cloud.


\subsubsection{Comparative Assessment of the Proposed Scenarios}
The three mechanisms considered above---leptonic bremsstrahlung, hadronic proton-gas interactions, and accretion onto a compact object---differ
substantially in their physical requirements and energetic feasibility.  Table \ref{tab:model_matrix} summarizes their key properties.

The leptonic scenario is strongly disfavored because reproducing the observed GeV luminosity requires an electron power far exceeding the mechanical output of SS 433.  The hadronic mechanism is less demanding energetically, but still requires either efficient confinement or unusually favorable transport of relativistic protons into the cloud.  The accretion scenario, in contrast, satisfies the basic energetic requirements under reasonable physical conditions for a compact object
embedded in the molecular gas.

At present, none of these scenarios can be conclusively ruled out, but the balance of energetics and environmental constraints indicates that the accretion interpretation is the most self-consistent among the three.  Further observational tests will be required to distinguish these possibilities.

\begin{figure*}[hbtp]
\begin{center}
\includegraphics[clip, width=0.6\textwidth]{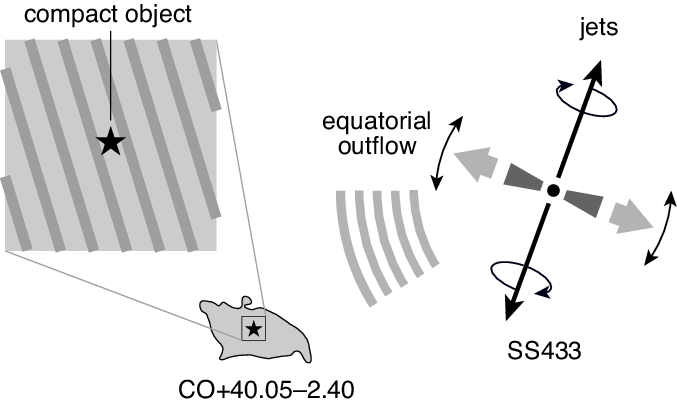}
\caption{Schematic illustration of the mechanism for producing periodic variability in the accretion scenario.} 
\label{fig:heartbeat}
\end{center}
\end{figure*}

\subsection{Scenarios for the Gamma-ray Periodicity} \label{subsec:periodicity}
A defining feature of Fermi J1913+0515 is its periodic or quasi-periodic $\gamma$-ray modulation, sometimes referred to as a ``heartbeat.''  The reported period matches, to within uncertainties, the $\sim\!162$-day precession period of the SS 433 jets, suggesting a physical connection.  Such behavior is difficult to reproduce in models based solely on steady irradiation by cosmic rays.

In the accretion framework, variability arises if the mass supply is periodically modulated.  SS 433 is known to drive a precessing equatorial outflow in addition to its jets \citep{Fabrika_2004}, and such a flow interacting with the surrounding medium may generate compressive waves or quasi-periodic density structures.  If such perturbations propagate into CO+40.05--2.40 and pass over an embedded compact object, the Bondi accretion rate—and thus $L_\gamma$—would vary in phase with the SS 433 precession.

In this interpretation, the observed $\gamma$-ray periodicity reflects modulation of the inflowing gas rather than modulation of particle acceleration.  This scenario provides a unified explanation for both the luminosity and variability without requiring unrealistically large cosmic-ray fluxes.

While speculative, the model makes testable predictions including:  
(1) substructure or velocity asymmetry in the cloud consistent with propagating density waves;  
(2) phase-resolved spectral changes in the $\gamma$-ray emission; and  
(3) the presence of a faint compact X-ray or radio source near the cloud center.  
Future molecular-line mapping and multi-wavelength monitoring will be essential to evaluate whether the periodicity indeed reflects accretion regulated by the precessing SS 433 outflow.

\begin{table*}[htbp]
\centering
\caption{Comparison of the Proposed Gamma-ray Emission Scenarios}
\label{tab:model_matrix}
\begin{tabular}{lccc}
\hline\hline
Category &
Leptonic &
Hadronic &
Accretion \\
\hline
Energy source &
SS\,433 $e^{-}$ &
SS\,433 $p$ &
Bondi accretion \\
Required power &
$\gg L_{\rm mech}^{\Omega}$ &
$> L_{\rm mech}^{\Omega}$ &
Matches $L_{\gamma}$ \\
Key requirement &
Extreme $e^{-}$ flux &
Confinement/transport &
Embedded compact object \\
Geometric constraint &
Strong (small $\Omega$) &
Strong (small $\Omega$) &
None \\
Cloud condition &
No heating/shock &
No heating/shock &
Density sets $\dot{M}$ \\
162-d periodicity &
No &
No &
Possible \\
Observational tests &
$e^{-}$ synchrotron? &
$\pi^{0}$ bump? &
Compact counterpart? \\
Overall viability &
Low &
Moderate &
High \\
\hline
\end{tabular}
\end{table*}

\section{Conclusions} \label{sec:conclusions}
We have identified a molecular cloud, CO+40.05--2.40, lying within the 95\% positional uncertainty region of the GeV source Fermi J1913+0515 at the northern boundary of the SS 433/W50 system. Millimeter and submillimeter observations show that the cloud exhibits physical and chemical properties typical of quiescent Galactic disk molecular gas, with no indications of shock heating, external compression, or enhanced excitation. A statistical analysis of CO survey data demonstrates that its spatial coincidence with Fermi J1913+0515 is unlikely to be accidental, supporting a physical association.

We evaluated three plausible $\gamma$-ray emission mechanisms.
(1) Leptonic bremsstrahlung is strongly disfavored: the required electron power far exceeds the geometrically available energy from SS 433.
(2) Hadronic proton-gas interactions are energetically less demanding but still require proton injection rates larger than the power that SS 433 can supply unless particle transport is restricted to a very narrow solid angle toward the cloud.
(3) Accretion onto an embedded compact object can naturally supply the required luminosity through Bondi accretion under reasonable physical conditions in the cloud. This scenario also provides a framework in which the reported $\sim\! 162$-day modulation could arise, although such variability is not adopted here as a decisive diagnostic.

Further observations are needed to test these possibilities. High-resolution molecular-line imaging may reveal signatures of interaction with the SS 433 outflow or internal density structures, while deep X-ray and radio searches may detect a compact object if present. Continued $\gamma$-ray monitoring will clarify the stability and spectral behavior of the emission.  These efforts will determine whether CO+40.05--2.40 hosts an as-yet unrecognized compact $\gamma$-ray source or represents a previously unrecognized class of particle-cloud interaction associated with SS 433.

\begin{acknowledgments}
The results presented in this paper are based on data obtained using the Nobeyama Radio Observatory (NRO) and the James Clerk Maxwell Telescope (JCMT).  We thank Haruka Sakemi for providing the JVLA radio image and Naomi Tsuji for providing the Chandra X-ray image used in Figure 1.  The NRO is a branch of the National Astronomical Observatory of Japan, National Institutes of Natural Sciences.  The JCMT is operated by the East Asian Observatory on behalf of the National Astronomical Observatory of Japan; Academia Sinica Institute of Astronomy and Astrophysics; the Korea Astronomy and Space Science Institute; the National Astronomical Research Institute of Thailand; Center for Astronomical Mega-Science (as well as the National Key R\&D Program of China with No. 2017YFA0402700).  Additional funding support is provided by the Science and Technology Facilities Council of the United Kingdom and participating universities and organizations in the United Kingdom and Canada.  

We are grateful to the NRO staff for facilitating the operation of the telescope.  T.O. acknowledges the support from JSPS Grant-in-Aid for Scientific Research (A) No. 20H00178. 
\end{acknowledgments}

\bibliographystyle{aasjournal}

\end{document}